\documentclass[a4paper,nofootinbib]{revtex4} 
\usepackage{slashed,amsmath,amssymb,wrapfig} 
\usepackage{color,epsfig}
\usepackage{hyperref}

\newcommand{\beq} {\begin{equation}}
\newcommand{\eeq} {\end{equation}}
\newcommand{\beqa} {\begin{eqnarray}}
\newcommand{\eeqa} {\end{eqnarray}}

\newcommand{\ie}{{\it i.e.}}

\newcommand{\as}{{\alpha_s}}
\newcommand{\lqcd}{\Lambda_{QCD}}

\newcommand{\ieps}{i\varepsilon}

\newcommand{\order}[1]{${\cal O}\left(#1 \right)$}
\newcommand{\morder}[1]{{\cal O}\left(#1 \right)}
\newcommand{\eq}[1]{(\ref{#1})}
\newcommand{\fig}[1]{Fig.~\ref{#1}}

\newcommand{\inv}[1]{\frac{1}{#1}}
\newcommand{\halft}{{\textstyle \frac{1}{2}}}

\newcommand{\quart}{{\textstyle \frac{1}{4}}}

\newcommand{\ket}[1]{\left\vert{#1}\right\rangle}
\newcommand{\bra}[1]{\langle{#1}\vert}

\newcommand{\com}[2]{\left[{#1},{#2}\right]}

\newcommand{\acom}[2]{\left\{{#1},{#2}\right\}}

\newcommand{\tr}{\mathrm{Tr}\,}

\newcommand{\bs}[1]{\boldsymbol{#1}}

\newcommand{\xv}{{\bs{x}}}

\newcommand{\kv}{{\bs{k}}}

\newcommand{\Pv}{{\bs{P}}}

\newcommand{\gv}{\bs{\gamma}}
\newcommand{\delv}{\bs{\delta}}
\newcommand{\gz}{\gamma^0}

\newcommand{\nv}{\bs{\nabla}}

\newcommand{\xbj}{{x_{Bj}}}

\begin{document}

\title{Confinement with Perturbation Theory, after All?\footnote{Based on talks at {\it Light Cone 2014} (Raleigh, NC USA, May 2014) and at the {\it FAIR Workshop} (Kolymbari, Greece, July 2014).}}

\author{Paul Hoyer}
\affiliation{Department of Physics\\ POB 64, FIN-00014 University of Helsinki, Finland}

\begin{abstract} 
\noindent
I call attention to the possibility that QCD bound states (hadrons) could be derived using rigorous Hamiltonian, perturbative methods. Solving Gauss' law for $A^0$ with a non-vanishing boundary condition at spatial infinity gives an \order{\alpha_s^0} linear potential for color singlet $q\bar q$ and $qqq$ states. These states are Poincar\'e and gauge covariant and thus can serve as initial states of a perturbative expansion, replacing the conventional free $in$ and $out$ states. The coupling freezes at $\alpha_s(0)\simeq 0.5$, allowing reasonable convergence. The \order{\alpha_s^0} bound states have a sea of $q\bar q$ pairs, while transverse gluons contribute only at \order{\as}. Pair creation in the linear $A^0$ potential leads to string breaking and hadron loop corrections. These corrections give finite widths to excited states, as required by unitarity. Several of these features have been verified analytically in $D=1+1$ dimensions, and some in $D=3+1$.

\end{abstract}


\maketitle

\vspace{-.5cm}


\parindent 0cm
\vspace{-.2cm}

\subsection*{1. Hadron physics}

Presently, numerical lattice methods are our only {\it first principles} approach to hadron physics \cite{Beringer:1900zz}. Lattice calculations demonstrate that QCD describes soft (confinement) physics and give valuable information on hadron spectra and structure. There are good reasons to believe that analytic, perturbative bound state methods (which work for QED atoms) are inapplicable to hadrons. To name a few of the arguments:
\begin{enumerate}

\item[I.] {\it Confinement} does not arise from Feynman diagrams at any finite order in $\as$.

\item[II.] {\it Chiral symmetry} is preserved at each order of perturbation theory (for $m_q \to 0$).

\item[III.] Hadron wave functions have abundant {\it sea quark and gluon} constituents.

\end{enumerate}
In the absence of an analytic method based directly on the QCD action other approaches to hadron dynamics have been developed. These include expansions based on kinematic limits (twist expansion, chiral perturbation theory, heavy quark effective theory). A variety of models involving some ad hoc assumptions (quark models, Dyson-Schwinger approaches,\ldots) have also provided insights. 

Meanwhile, the principles of perturbative bound state calculations are largely ignored in modern courses on field theory. This is unwarranted:
\begin{itemize}

\item Bound states provide insights to perturbation theory which are complementary to those of scattering phenomena.

\item Convincing as the above arguments I, II, III may seem, there is a risk that we are ``throwing out the baby with the bathwater'' \cite{Dokshitzer:1998nz}.

\end{itemize}

Hadron data has features which point to a perturbative context. Quarkonium spectra have a strikingly atomic appearance (``The $J/\psi$ is the Hydrogen atom of QCD''). The valence quark degrees of freedom (but not those of sea quarks and gluons) are manifest also in light hadron spectra. ``OZI forbidden'' decays such as $\phi \to \pi\pi\pi$ are suppressed, even though they can proceed via soft gluon exchanges.

In the following I summarize a search for a perturbative approach to soft QCD processes which is compatible with basic facts about hadrons, including the points mentioned above. The requirements of theoretical consistency strongly constrain the \order{\alpha_s^0} bound states which may serve as initial states in a perturbative expansion. I refer to published work \cite{Hoyer:1986ei} and lecture notes \cite{Hoyer:2014gna} for details. Naturally, further work may uncover inconsistencies or poor quantitative agreement with data.

\subsection*{2. The strong coupling}

Data on hard processes has verified the running of $\as(Q)$ as predicted by QCD at high scales $Q$. The coupling reaches $\as \simeq 0.33$ at $Q=m_\tau = 1.8$ GeV \cite{Beringer:1900zz}. There are indications that the running stops at hadronic scales, with the coupling freezing at a value $\as(Q=0) \simeq 0.5$ \cite{Dokshitzer:1998qp}.

Perturbative gluons are absent at \order{\alpha_s^0} in the present scenario. Consequently little running is expected until scales where radiative gluon effects become important. We may {\it very roughly} estimate this scale by assuming that $\as(Q)$ freezes suddenly at $Q=Q_0$. Setting $\as(Q_0)=0.5$ in the standard perturbative expression ($n_f=3$, LO) with $\lqcd=200$ MeV gives $Q_0 \simeq 800$ MeV.

The magnitude of the QCD coupling could bring about a fundamental change in the structure of the vacuum, as observed by Gribov \cite{Gribov:1999ui}. He noted that there is a critical coupling $\alpha_s^{crit}\simeq 0.43$ at which the Coulomb attraction between light fermions becomes strong enough to make the pair energy negative. In this scenario the running of $\as(Q)$ triggers confinement at $Q \simeq 1$ GeV.

\subsection*{3. Positronium}

Bound state poles of QED scattering amplitudes arise from the {\it divergence} of the perturbative expansion. No finite order Feynman diagram for $e^+e^- \to e^+e^-$ can have a pole at the Positronium (rest frame) energy $E_0=2m_e-\quart \alpha^2 m_e$, since $1/(E-E_0)$ is non-polynomial in $\alpha$. The Feynman diagrams which contribute to the divergence at leading order may be identified as single photon ladder diagrams. Their sum implies the familiar Schr\"odinger equation for the residue of the pole (\ie, for the wave function).

The divergence of the perturbative expansion is caused by our choice of \order{\alpha^0} initial states. The free $in$ and $out$ electron states are stripped of their electromagnetic fields and thus do not satisfy the field equations of motion. The EM fields are restored by the sum of ladder diagrams, enabling bound states. It is no accident that the ladder sum generates precisely the {\it classical} $eA^0 = -\alpha/r$ potential, which satisfies the field equations. This may be viewed as the Born approximation for bound states, analogous to tree diagrams for scattering amplitudes. Loop corrections to Born bound states give higher order corrections to $E_0$, such as the Lamb shift.

The Schr\"odinger equation may be derived more directly from the QED action by starting from a general $e^+e^-$ state in its rest frame, defined by a product of normal ordered fields,
\beq\label{posstate}
\ket{e^-e^+,t} = \int d^3\xv_1\,d^3\xv_2\,\bar\psi_\alpha(t,\xv_1)\Phi_{\alpha\beta}(\xv_1-\xv_2)\psi_{\beta}(t,\xv_2)\ket{0}
\eeq
The $\bar\psi$ field creates an electron at $\xv_1$ and $\psi$ a positron at $\xv_2$, at the common time $t$. The $c$-numbered wave function $\Phi_{\alpha\beta}(\xv_1-\xv_2)$ is a $4\times 4$ matrix in Dirac indices and determines the distribution of the pair in space. For \eq{posstate} to represent a bound state it must be stationary in time,
\beq\label{bscond}
H_{QED}(t)\ket{e^-e^+,t} = E\ket{e^-e^+,t}
\eeq
where the QED Hamiltonian is derived from the action in the usual way \cite{Weinberg:1995mt}. The Born approximation implies replacing the photon field operator $A^\mu(t,\xv)$ in $H_{QED}$ by the classical potential. For non-relativistic positronium at rest the Coulomb potential $A^0$ dominates. With an electron at $\xv_1$ and a positron at $\xv_2$ it satisfies the field equation (Gauss' law),
\beqa
-\nv^2 A^0(t,\xv)&=&e\big[\delta^3(\xv-\xv_1)-\delta^3(\xv-\xv_2)\big] \label{gausslaw} \\[2mm]
A^0(t,\xv)&=&\frac{e}{4\pi}\left(\inv{|\xv-\xv_1|}-\inv{|\xv-\xv_2|}\right) \label{coulsol}
\eeqa
Using this in $H_{QED}$ ({\it separately} for each component $\ket{e^-(\xv_1)e^+(\xv_2),t}$ of the bound state), and neglecting pair production, the eigenvalue equation \eq{bscond} gives the bound state equation for the wave function $\Phi$,
\beq\label{bse}
i\nv\cdot\acom{\gamma^0\gv}{\Phi(\xv)}+m\com{\gamma^0}{\Phi(\xv)} = \big[E-V(\xv)\big]\Phi(\xv)
\eeq
Here $\xv \equiv \xv_1-\xv_2$ and $V(\xv)=\halft \big[eA^0(t,\xv_1)-eA^0(t,\xv_2)\big]=-\alpha/|\xv|$ (discarding the infinite constant $\alpha/0$). Eq. \eq{bse} is valid in the non-relativistic limit, where it reduces to the Schr\"odinger equation.

\subsection*{4. The linear potential}

Relativistic bound states may be derived from the QCD action using a method similar to the above one for Positronium. For conciseness I consider only $q\bar q$ states and abelian U(1) gauge invariance. The generalization to SU(3) of color does not bring anything conceptually new, apart from $qqq$ solutions.

Bound state poles in QCD scattering amplitudes arise (as in QED) from a divergence of the sum of Feynman diagrams. However, we do not know which diagrams to sum for a first approximation -- ladder diagrams dominate only for non-relativistic states. Instead we may turn the question around and ask what classical gluon potential the sum can possibly generate, given that it should satisfy the equations of motion and maintain Poincar\'e invariance. The Born approximation is a fully relativistic concept.

There is no $\partial_0 A^0$ term in gauge theory Lagrangians. Gauss' law \eq{gausslaw} determines $A^0$ for each charge configuration at each instant of time. The QED solution \eq{coulsol} for $A^0$ is obtained assuming $A^0(|\xv|\to\infty)=0$. The simplest homogeneous solution with a non-vanishing field at spatial infinity is linear in $\xv$,
\beq\label{linsol1}
A^0(t,\xv) =\kappa\, \xv\cdot(\xv_1-\xv_2) + \morder{g}
\eeq
The dot product with $\xv_1-\xv_2$ is imposed by rotational invariance and $\kappa$ may depend on $\xv_1,\xv_2$. The square of the field strength density
\beq
\big[\nv A^0\big]^2 = \kappa^2\,(\xv_1-\xv_2)^2 + \morder{g}
\eeq
contributes a divergent term $\propto \int d^3\xv$ to the field energy. This is irrelevant provided it is independent of the quark positions ($\xv_1,\xv_2$). Hence we must have $\kappa=\Lambda^2/|\xv_1-\xv_2|$, where $\Lambda$ is a universal constant with dimension of energy. The bound state potential is consequently linear,
\beq\label{linpol}
V(\xv)=\halft g\big[A^0(t,\xv_1)-A^0(t,\xv_2)\big]=\halft g\Lambda^2\,|\xv_1-\xv_2| + \morder{g^2}
\eeq
Note that the potential is invariant under translations $(\xv_1,\xv_2)\to (\xv_1+\bs{a},\xv_2+\bs{a})$ only for {\it neutral} states. For SU(3) of color translation invariance similarly requires {\it color singlet} states. 

The linear solution \eq{linsol1} is the only homogeneous solution of Gauss' law that preserves Poincar\'e symmetry. Quadratic or higher powers of $\xv$ break translation invariance. Boost covariance also requires a linear potential (Sect. 6.1).
The \order{g^2} potential in \eq{linpol} includes the perturbative (abelian) gluon exchange contribution $-g^2/4\pi|\xv_1-\xv_2|$ as well as loop contributions of higher order in $\hbar$.

The constant $\Lambda$ of the linear potential in (8) determines the radius of the bound states, and hence also the scale at which soft gluons decouple from color singlet hadrons. This regulates the infrared singularities of the perturbative expansion, and sets the scale $Q_0$ at which the coupling freezes. The scale which determines $\alpha_s(Q)$ at high $Q$ will be $\propto\Lambda$, with the proportionality constant dependent on the value chosen for the frozen coupling $\alpha_s(0)=g^2/4\pi$. Requiring this scale to be $\Lambda_{QCD}\simeq 200$~MeV fixes the value of $g$. A detailed study of the perturbative corrections is, however, worthwhile only provided the \order{\alpha_s^0}, or strictly speaking \order{g}, bound states turn out to be viable as asymptotic states in a perturbative expansion, in place of the free $in$ and $out$ quark and gluon states.

In the bound state condition \eq{bscond} for non-relativistic Positronium we neglected pair production in the vacuum,
\beq\label{nopair}
A^0(\xv)\psi^\dag(t,\xv)\psi(t,\xv)\ket{0} \to 0
\eeq
With relativistic dynamics pair production occurs when $A^0\neq 0$. Specifically, in a $\ket{q(\xv_1)\bar q(\xv_2)}$ state the linear potential \eq{linpol} causes pair creation (string breaking). This implies decay and loop corrections to the Born states determined by \eq{bse} (Sect. 6.2).

\subsection*{5. Multiparticle nature of the Dirac wave function}

The bound state equation \eq{bse} resembles a double Dirac equation \cite{Breit:1929zz}. It is instructive to consider how the usual Dirac equation emerges in the present framework. Let the state 
\beq\label{diracstate}
\ket{e^-,t}=\int d^3\xv\,\psi_\alpha^\dag(t,\xv)\Psi_\alpha(\xv)\ket{0}_R
\eeq
represent an electron bound in a static external potential $A_{ext}^0(\xv)$, with $\Psi_\alpha(\xv)$ its $c$-numbered Dirac wave function.
The eigenvalue condition $H_{QED}(A_{ext}^0)\ket{e^-,t}=E\ket{e^-,t}$ gives the Dirac equation for $\Psi(\xv)$,
\beq\label{direq}
(-i\nv\cdot\gamma^0\gv+m\gamma^0)\Psi(\xv)=(E-eA^0_{ext})\Psi(\xv)
\eeq
{\it provided} we neglect pair production as in \eq{nopair}. However, this is unjustified for $A^0=A^0_{ext} \neq 0$ when the electron is relativistic. Pair creation indeed occurs: The Dirac state is a superposition of Fock states with any number of $e^+e^-$ pairs (as demonstrated by the solution of the Klein paradox \cite{Hansen:1980nc}). But what is then the meaning of the ``single electron'' Dirac wave function $\Psi(\xv)$ which solves \eq{direq}?

An inspection of the Feynman diagrams that describe the electron scattering in $A^0_{ext}$ shows that identical energy eigenvalues $E$ are obtained using Feynman and retarded electron propagators. The electron energy $p^0$ is constant since the static potential only transfers 3-momentum $\kv$. For $p^0>0$ the $\ieps$ prescription at the negative energy pole ($p^0=-E$) of the electron propagator is irrelevant\footnote{This argument breaks down when $A^0_{ext}$ is sufficiently strong to make the bound state energy negative.}. In retarded propagation the negative energy electrons move forward in time, which eliminates the pair-creating $Z$-diagrams of Feynman propagation. The Dirac wave function $\Psi(\xv)$ in \eq{direq} then describes the single (positive or negative energy) electron which with retarded boundary conditions is present at any intermediate time.

In the definition \eq{diracstate} of the Dirac state the retarded boundary condition is indicated by $\ket{0}_R$. A retarded propagator $S_R(x-y)={}_R\bra{0}{\rm T}\big[\psi(x)\bar\psi(y)\big]\ket{0}_R \propto \theta(x^0-y^0)$ requires $\psi(x)\ket{0}_R=0$. This validates \eq{nopair} with $\ket{0}\to \ket{0}_R$. Due to the Pauli exclusion principle the retarded fermion vacuum may be expressed as
\beq\label{retvac}
\ket{0}_R= \prod_{\kv,\lambda}d^\dag_{\kv,\lambda}\ket{0}
\eeq
where the product is over all momenta $\kv$ and helicities $\lambda$. Thus both the $b^\dag$ and $d$ operators in the $\psi^\dag$ field of \eq{diracstate} contribute, with $d$ creating negative energy states through the removal of a positive energy $d^\dag$ from $\ket{0}_R$.

The retarded boundary condition gives the Dirac wave function an ``inclusive'' character. The operator
\beq\label{denseop}
\int d^3\xv\,\psi^\dag(t,\xv)\psi(t,\xv) = \int \frac{d^3\kv}{(2\pi)^3}\sum_{\lambda}\, (b^\dag_{\kv,\lambda} b_{\kv,\lambda}+d_{\kv,\lambda} d^\dag_{\kv,\lambda})
\eeq
is normally interpreted as the charge operator 
due to the reordering $d d^\dag\to -d^\dag d$. In the retarded vacuum \eq{retvac} $d^\dag$ is the annihilation operator and thus \eq{denseop} is the {\it number} operator. The expectation value
\beq\label{dirdensity}
\bra{e^-,t}\psi_\alpha^\dag(t,\xv)\psi_\alpha(t,\xv)\ket{e^-,t} = \Psi^\dag(\xv)\Psi(\xv)\ {_{R}\bra{0}}0\rangle_{R}
\eeq
shows that $|\Psi(\xv)|^2$ is the density of positive and negative energy electrons in the Dirac state. Thus the norm of the Dirac wave function should be interpreted as an {\it inclusive} particle density.

\subsection*{6. Properties of the $q\bar q$ bound states}

{\it 6.1 Boost covariance}  

Bound states must transform covariantly under boosts to ensure the Poincar\'e invariance of their matrix elements. In a frame where the momentum of the $q\bar q$ state is $\Pv$ Eqs. \eq{posstate} and \eq{bse} become
\beqa
\ket{E,\Pv,t} = \int d^3\xv_1\,d^3\xv_2\,\bar\psi(t,\xv_1)\exp\big[i\halft \Pv\cdot(\xv_1+\xv_2)\big]\Phi(\xv_1-\xv_2)\psi(t,\xv_2)\ket{0}\label{qqstate} \\[2mm]
i\nv\cdot\acom{\gamma^0\gv}{\Phi(\xv)}-\halft \Pv\cdot\com{\gamma^0\gv}{\Phi(\xv)}+m\com{\gamma^0}{\Phi(\xv)} = \big[E-V(\xv)\big]\Phi(\xv) \label{qqwf}
\eeqa 
with $V(\xv)$ as in \eq{linpol}. States defined at equal time transform dynamically under boosts. There is no previous experience for how the wave function $\Phi(\xv)$ should depend on $\Pv$, but we know its eigenvalue $E(\Pv)=\sqrt{M^2+\Pv^2}$.

In $D=1+1$ dimensions the bound state equation \eq{qqwf} reduces to two differential equations coupling the components $\phi_0$ and $\phi_1$ of the $2\times 2$ wave function $\Phi=\phi_0+\sigma_1\phi_1+\sigma_3\phi_2 
+i\sigma_2\phi_3$,
\beq\label{eom3}
-2i\frac{\partial\phi_1}{\partial\sigma}=\phi_0 \hspace{1cm} {\rm and} \hspace{1cm} -2i\frac{\partial\phi_0}{\partial\sigma}=\Big(1-\frac{4m^2}{\sigma}\Big)\phi_1 \hspace{1cm} {\rm where}\ \ \ \sigma(x)\equiv [E-V(x)]^2-P^2
\eeq
The differential equations have no explicit $P$-dependence when the potential $V(x)$ in \eq{qqwf} is linear, making $\phi_{0,1}(\sigma)$ frame independent. The $P$-dependence of the function $\sigma(x)$ in \eq{eom3} determines the frame dependence of $\Phi$ when viewed as a function of $x$. $\Phi(x)$ is regular at $\sigma=0$ only for discrete energy eigenvalues $E(P)$, which turn out to have the $P$-dependence required by Lorentz invariance.

The fact that the energy eigenvalues of the bound state equation \eq{qqwf} are boost covariant is a necessary but not sufficient condition for the {\it states} to transform according to the boost operator $U$ determined by the action,
\beq\label{boostop}
U(\xi)\ket{E,\Pv,t}=\ket{E',\Pv',t}
\eeq
where $(E,\Pv)$ transforms into $(E',\Pv')$ in the boost characterized by $\xi$. The boost property \eq{boostop} of the states was shown to hold in $D=1+1$, and also in higher dimensions for the ``collinear'' configuration $\xv_1-\xv_2\parallel \Pv$ in \eq{qqstate}. This boost covariance holds only for linear potentials (in any dimension). The covariance \eq{boostop} remains to be demonstrated for general $\xv_1-\xv_2$.

{\it 6.2 Normalization}

The normalization integral of the Dirac wave function diverges for a linear $A^0$ potential \cite{plesset}. In $D=1+1$ $\Psi(x)$ is given by Confluent Hypergeometric functions and it is readily seen that $|\Psi(x)|^2$ approaches a constant at large $x$. According to \eq{dirdensity} this means that the linear potential creates a constant density of virtual $q\bar q$ pairs. Numerically one may verify that $\Psi(x)$ approximates the exponentially decreasing Airy function of the Schr\"odinger equation in the non-relativistic region $V(x) \ll m$, but then increases again at distances $x$ where $V(x) \simeq 2m$.

The norm of the $q\bar q$ wave function $\Phi$ of \eq{qqwf} similarly approaches a constant at large $|\xv|$ \cite{Geffen:1977bh}. Since the bound state equation neglects string breaking the interpretation is analogous: $|\Phi(\xv)|^2$ gives the {\it inclusive} distribution of $q$ and $\bar q$ in the state. Pair production/annihilation may now be included iteratively. If $A,B$ and $C$ are states of the form \eq{qqstate}, the $A\to B+C$ matrix element is found by contracting a field operator in $B$ with one in $C$, giving
\beqa\label{hadvert}
\bra{B,C}A\rangle =-\frac{(2\pi)^3}{\sqrt{N_C}}\delta^3(\Pv_A-\Pv_B-\Pv_C)\int d\delv_1 d\delv_2\,e^{i\delv_1\cdot\Pv_C/2-i\delv_2\cdot\Pv_B/2}\tr\big[\gz\Phi_B^\dag(\delv_1)\Phi_A(\delv_1+\delv_2)\Phi_C^\dag(\delv_2)\big]
\eeqa
%
\begin{wrapfigure}[15]{r}{0.3\textwidth}
  \vspace{-25pt}
  \begin{center}\hspace{-.5cm}
    \includegraphics[width=0.25\textwidth]{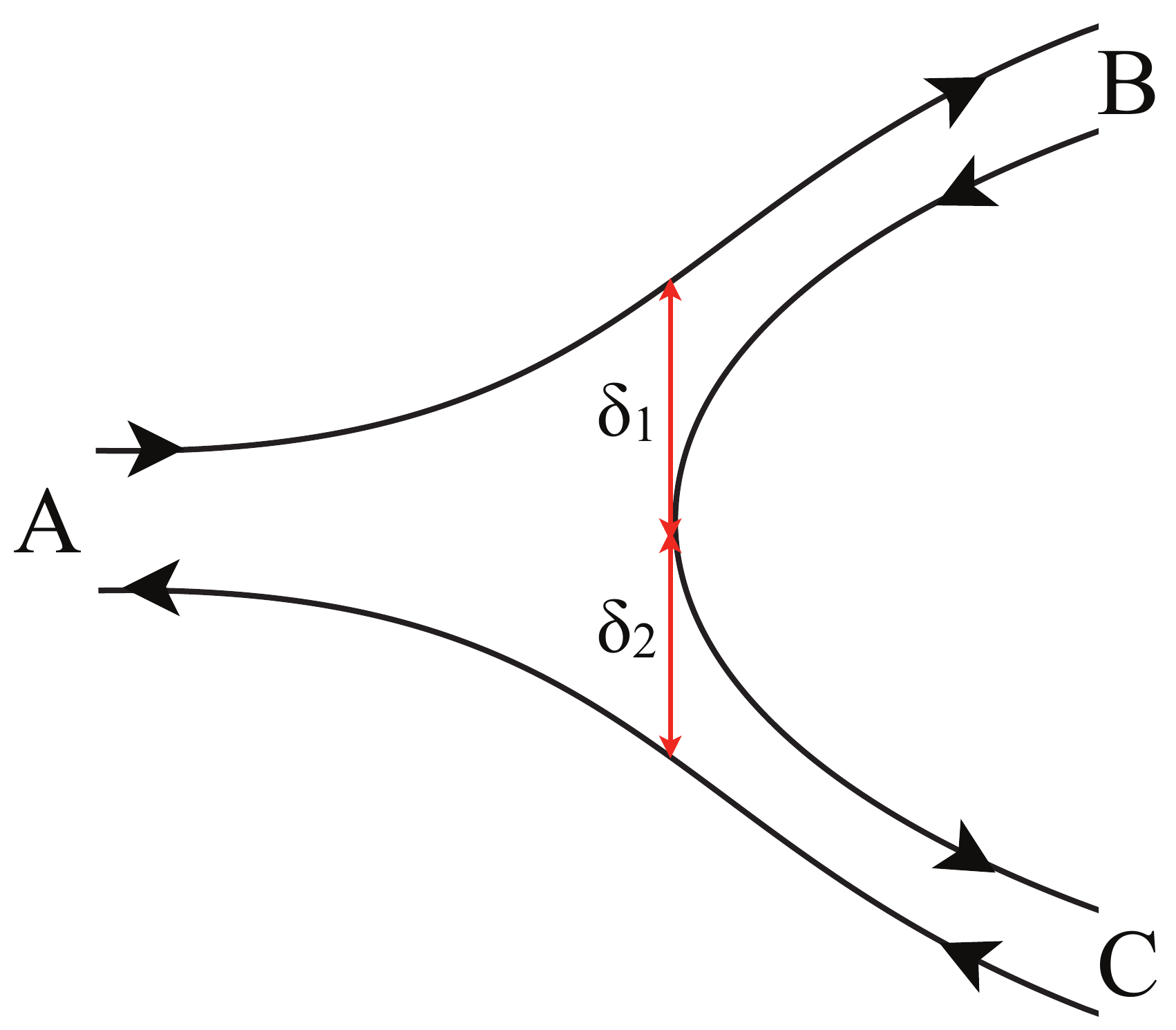}
  \end{center}
  \vspace{-15pt}
  \caption{The diagram for meson decay $A\to B+C$, given by \eq{hadvert}. The $q\bar q$ pair is created at distance $\delv_1$ from the quark and $\delv_2$ from the antiquark of $A$.}\label{dualdiag}
\end{wrapfigure}
%
where $N_C$ is the number of colors. As seen in \fig{dualdiag} this matrix element resembles a dual diagram. The boost covariance \eq{boostop} ensures its Poincar\'e invariance, despite appearances. It remains to be demonstrated that such ``unitarity corrections'' associate the large $|\xv|$ components of $\Phi(\xv)$ with multimeson final states, leaving single hadrons with normalizable wave functions.

{\it 6.3 Absence of parity doublets for any $m\neq 0$}

The bound state equation \eq{qqwf} reflects the chiral symmetry of the action for $m=0$. If $\Phi(\xv)$ is a solution then so are $\gamma_5\Phi(\xv)$ and $\Phi(\xv)\gamma_5$, with the same $E$. The $m=0$ states are thus parity degenerate. Nevertheless, the $m \neq 0$ solutions are {\it not} parity degenerate even in the limit $m\to 0$. As mentioned above, the term $m^2/\sigma$ in \eq{eom3} generally makes $\Phi(\xv)$ singular at $\sigma=0$. The requirement of (local) normalizability at $\sigma=0$ gives a discrete energy spectrum, which is not parity doubled for any finite $m$. For $m=0$ the $\sigma=0$ singularity is absent, implying a continuous (in particular, parity degenerate) spectrum. 

%
\begin{wrapfigure}[11]{r}{0.3\textwidth}
  \vspace{-30pt}
  \begin{center}\hspace{-.5cm}
    \includegraphics[width=0.3\textwidth]{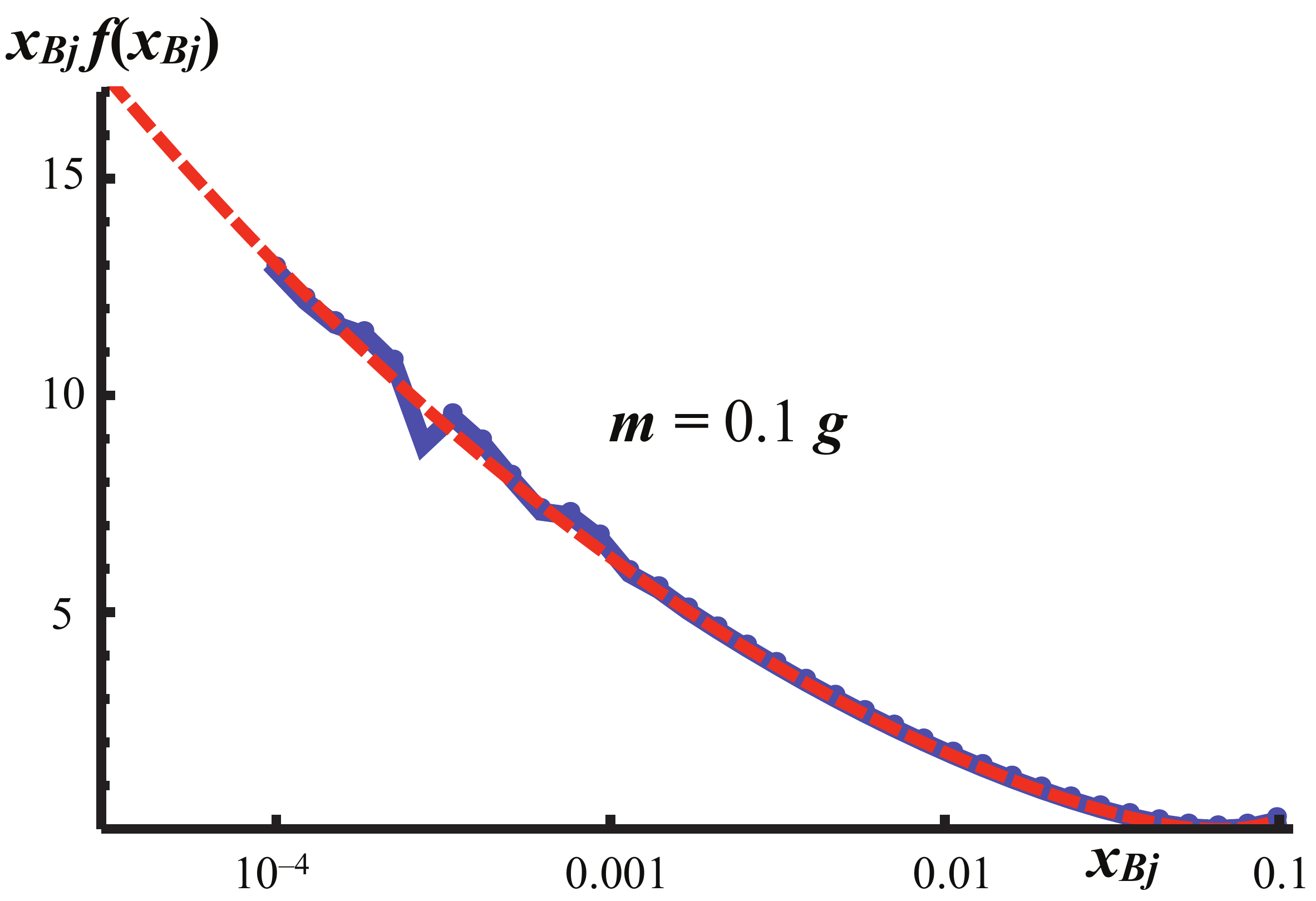}
  \end{center}
  \vspace{-15pt}
  \caption{The parton distribution of the ground state in $D=1+1$, numerically evaluated (blue dots) for fermion mass $m=0.1\, g$ in the region $\xbj<0.1$. The dashed red curve is an analytic approximation valid at small $\xbj$.}\label{partondist}
\end{wrapfigure}
%

In contrast, the Dirac wave function in \eq{direq} is regular at all $\xv$ and so the Dirac energy spectrum is continuous for {\it any} $m$ when the potential $eA^0_{ext}$ is linear \cite{titchmarsh}.

{\it 6.4 Electromagnetic form factors and DIS}

Electromagnetic form factors may be defined using the bound states \eq{qqstate} as asymptotic $(t=\pm\infty)$ states,
\beq\label{formfac}
F^\mu_{AB}(z) = \bra{B(P_B);t=+\infty}j^\mu(z)\ket{A(P_A);t=-\infty}
\eeq
where $j^\mu(z)= \bar\psi(z)\gamma^\mu\psi(z)$. The bound state equation \eq{qqwf} for $\Phi_{A,B}$ ensures gauge invariance, $\partial_\mu F^\mu_{AB}(z)=0$. 

The transition form factors $\gamma^* A\to B$ describe (via duality) Deep Inelastic Scattering on target $A$, with $M_B \to\infty$ in the Bj limit. The quark distribution evaluated in $D=1+1$ dimensions is shown in \fig{partondist} for $\xbj<0.1$. The increase for $\xbj\to 0$ indicates contributions from the $q\bar q$ pairs in the target state.

\subsection*{7. Concluding remarks}

First principles (Hamiltonian) approaches to QCD in the confinement regime deserve attention, notwithstanding challenges like I, II, III recalled in Sect.~1. Striking features of the data (hadron spectra reflecting their valence quark degrees of freedom, ``atomic'' quarkonium spectra, OZI rule, duality, \ldots) indicate a perturbative formulation. The approach summarized above gives tentative answers to the three challenges (in Sects. 4, 6.3 and 5, respectively). The startling feature of $q\bar q$ wave functions with constant norm at large distances (before string breaking)
enables duality between bound states and scattering as well as the parton picture.

Many aspects of the \order{\alpha_s^0} ``Born term'' amplitudes remain to be explored, not to speak of higher order corrections. The present approach will hopefully turn out to be relevant for hadron physics. It could also be useful for understanding general properties of relativistic bound states. For example, the boost covariance \eq{boostop} may reveal how angular momentum, which is well-defined in the rest frame, manifests itself in the infinite momentum frame \cite{Leader:2013jra}.

\acknowledgments
An important part of the work described here was done in collaboration with Dennis Dietrich and Matti J\"arvinen. I thank the organizers of Lightcone 2014 and the FAIR workshop for their invitation. Part of this work was done during a two month visit to the NIKHEF theory group. I have enjoyed a travel grant from the Magnus Ehrnrooth Foundation.


\vspace{1cm}

\begin{thebibliography}{999}

\bibitem{Beringer:1900zz}
  J.~Beringer {\it et al.}  [Particle Data Group],
  Phys.\ Rev.\ D {\bf 86} (2012) 010001.

\bibitem{Dokshitzer:1998nz}
  Y.~L.~Dokshitzer,
  In *Vancouver 1998, High energy physics, vol. 1* 305-324
  [hep-ph/9812252].

\bibitem{Hoyer:1986ei}
  P.~Hoyer,
  Phys.\ Lett.\  B {\bf 172} (1986) 101;\\
%
  P.~Hoyer,
  arXiv:0909.3045 [hep-ph];\\
%
  D.~D.~Dietrich, P.~Hoyer and M.~J\"arvinen,
  Phys.\ Rev.\ D {\bf 85} (2012) 105016
  [arXiv:1202.0826 [hep-ph]];\\
%
  D.~D.~Dietrich, P.~Hoyer and M.~J\"arvinen,
  Phys.\ Rev.\ D {\bf 87} (2013) 065021
  [arXiv:1212.4747 [hep-ph]].

\bibitem{Hoyer:2014gna}
  P.~Hoyer,
  arXiv:1402.5005 [hep-ph].

\bibitem{Dokshitzer:1998qp}
  Y.~L.~Dokshitzer, G.~Marchesini and G.~P.~Salam,
  Eur.\ Phys.\ J.\ direct C {\bf 1} (1999) 3
  [hep-ph/9812487];\\
%
  S.~J.~Brodsky, S.~Menke, C.~Merino and J.~Rathsman,
  Phys.\ Rev.\  D {\bf 67} (2003) 055008
  [arXiv:hep-ph/0212078];\\
%
  G.~Grunberg,
  Phys.\ Rev.\  D {\bf 73} (2006) 091901
  [arXiv:hep-ph/0603135];\\
%
  C.~S.~Fischer,
  J.\ Phys.\ G {\bf 32} (2006) R253
  [arXiv:hep-ph/0605173];\\
%
  A.~Deur, V.~Burkert, J.~P.~Chen and W.~Korsch,
  Phys.\ Lett.\  B {\bf 665} (2008) 349
  [arXiv:0803.4119 [hep-ph]];\\
%
  A.~C.~Aguilar, D.~Binosi, J.~Papavassiliou and J.~Rodriguez-Quintero,
  Phys.\ Rev.\ D {\bf 80} (2009) 085018
  [arXiv:0906.2633 [hep-ph]];\\
%
  T.~Gehrmann, M.~Jaquier, G.~Luisoni,
  Eur.\ Phys.\ J.\  {\bf C67 } (2010)  57-72.
  [arXiv:0911.2422 [hep-ph]];\\
%
  A.~Courtoy and S.~Liuti,
  Phys.\ Lett.\ B {\bf 726} (2013) 320
  [arXiv:1302.4439 [hep-ph]].
  
\bibitem{Gribov:1999ui}
  V.~N.~Gribov,
  Eur.\ Phys.\ J.\ C {\bf 10} (1999) 91
  [hep-ph/9902279];\\
%
  Y.~L.~Dokshitzer,
  hep-ph/0306287.  

\bibitem{Weinberg:1995mt}
  S.~Weinberg, Sec. 8.3 of
  ``The Quantum theory of fields. Vol. 1: Foundations,''
  Cambridge, UK: Univ. Pr. (1995).
  
\bibitem{Breit:1929zz}
  G.~Breit,
  Phys.\ Rev.\  {\bf 34} (1929) 553.

\bibitem{Hansen:1980nc}
  A.~Hansen and F.~Ravndal,
  Phys.\ Scripta {\bf 23} (1981) 1036.

\bibitem{plesset}
M. S. Plesset,
Phys.\ Rev. {\bf 41} (1932) 278.

\bibitem{Geffen:1977bh}
  D.~A.~Geffen and H.~Suura,
  Phys.\ Rev.\  D {\bf 16} (1977) 3305.

\bibitem{titchmarsh}
E. C. Titchmarsh,
Quart. J. Math. Oxford (2), 12 (1961), 227.

\bibitem{Leader:2013jra}
  E.~Leader and C.~Lorce,
  Phys.\ Rept.\  {\bf 541} 163
  [arXiv:1309.4235 [hep-ph]].

\end{thebibliography}
\end{document}